\documentclass[11pt]{article}
\usepackage{vietnam,epsfig}

\def\Journal#1#2#3#4{{#1} {\bf #2}, #3 (#4)}

\def\PRL{\em Phys. Rev. Lett.}
\def\PRD{{\em Phys. Rev.} D}

\def\ApJ{{\em Astrophys. J.} }
\def\RMP{{\em Rev. Mod. Phys.} }
\def\JHEP{{\em JHEP} }

\def\gtrsim{\mathrel{\mathchoice {\vcenter{\offinterlineskip\halign{\hfil
$\displaystyle##$\hfil\cr>\cr\sim\cr}}}
{\vcenter{\offinterlineskip\halign{\hfil$\textstyle##$\hfil\cr>\cr\sim\cr}}}
{\vcenter{\offinterlineskip\halign{\hfil$\scriptstyle##$\hfil\cr>\cr\sim\cr}}}
{\vcenter{\offinterlineskip\halign{\hfil$\scriptscriptstyle##$\hfil
\cr>\cr\sim\cr}}}}}
\def\lesssim{\mathrel{\mathchoice {\vcenter{\offinterlineskip\halign{\hfil
$\displaystyle##$\hfil\cr<\cr\sim\cr}}}
{\vcenter{\offinterlineskip\halign{\hfil$\textstyle##$\hfil\cr<\cr\sim\cr}}}
{\vcenter{\offinterlineskip\halign{\hfil$\scriptstyle##$\hfil\cr<\cr\sim\cr}}}
{\vcenter{\offinterlineskip\halign{\hfil$\scriptscriptstyle##$\hfil
\cr<\cr\sim\cr}}}}}

\begin{document}
\vspace*{4cm}
\title{Ultra-high energy cosmic ray propagation in the Universe}

\author{Martin Lemoine}

\address{Institut d'Astrophysique de Paris, C.N.R.S.,\\
 98 bis boulevard Arago, F-75014 Paris, France}

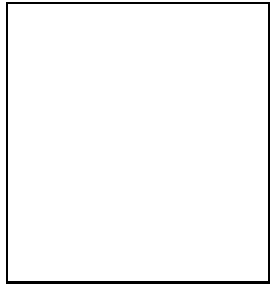
\maketitle\abstracts{This paper summarizes recent
developments in the study of the propagation of ultra-high energy
cosmic ray protons or nuclei in the Universe, with emphasis on the
consequences of energy losses and on the influence of extra-galactic 
magnetic fields. }

\section{Introduction}

Despite some forty years of intensive research, the origin and nature
of the highest energy cosmic rays ($E\sim 10^{20}\,$eV)  remains
elusive (see Ref.~\cite{NW00} for a review). 
On the experimental side, the matter is clearly one of
statistics: with a flux as low as $\sim 1/{\rm km}^2/{\rm century}$,
only a handful of events with $E\sim 10^{20}$eV have been gathered.
On the theoretical side, it appears notoriously difficult for known
astrophysical objects to produce particles of that energy. It is
furthermore problematic to explain how such particles could reach
the detectors with  energy $E\sim 10^{20}\,$eV, since the energy 
loss length $\sim 50\,$Mpc 
is very small by cosmological standards. Finally, there is no
observed counterpart in the arrival direction of these ultra-high
energy cosmic rays (UHECRs).

In the last decade, it has become apparent that propagation effects
could play a key r\^ole in the solution to this riddle. The aim of the present
paper is to present the salient features of these effects, with
particular emphasis on the influence of energy losses and of 
extra-galactic magnetic fields. The following discussion assumes
that UHECRs are protons or nuclei accelerated in an astrophysical 
source.

\section{What are UHECRs?}

Up to a few years ago it was generally admitted that the so-called
UHECR component emerged out of the all-particle cosmic ray spectrum at
the ``ankle'', a feature at $E\simeq 10^{19}\,$eV where the spectrum
index changes from $\approx-3.2$ to $\approx -2.7$, see~\cite{NW00}. 
A break associated to the
flattening of the spectrum is indeed characteristic of the emergence
of a new (harder) component. As protons with $E\gtrsim10^{19}\,$eV
cannot be confined in the Galactic magnetic field, it is assumed that
UHECRs are of extra-galactic origin. This interpretation agrees well
with the overall isotropicity of reported UHECRs arrival
directions~\cite{NW00}.

However, if the sources of UHECRs are located at cosmological
distances, one would expect to see a high-energy GZK
(Greisen-Zatsepin-Kuz'min) cut-off around
$\epsilon_{\rm GZK}\simeq6\cdot10^{19}\,$eV as a result of photomeson
production of UHE protons on the cosmic microwave background (CMB)~\cite{gzk}; 
a similar cut-off is expected for heavy nuclei due to
photodisintegration by the CMB~\cite{gzk-heavy}. The small energy loss
length implies that sources of particles with observed energy
$10^{20}$eV should lie closer than $\sim50\,$Mpc. At lower energies,
the energy loss length for protons increases to $\sim1\,$Gpc due to pair production
on the CMB and to $\sim4\,$Gpc due to redshift losses when $E\lesssim
10^{18.5}\,$eV.

 Present experimental data suggest the presence of some form of cut-off around
$\epsilon_{\rm GZK}$~\cite{BW}. 
However it is not clear at present how pronounced or how
mild this cut-off is~\cite{dMBO}. 
As a matter of act, the two experiments with the highest
statistics, namely HiRes~\cite{HiRes} and AGASA~\cite{AGASA}, disagree
with one another: 
while HiRes claims to detect a sharp cut-off at
$E\simeq\epsilon_{\rm GZK}$, the AGASA experiment has reported a
spectrum that extends in a featureless way beyond $10^{20}\,$eV, down
to the experimental sensitivity limit. These two experiments use
different techniques for reconstructing the primary energy, and one major
objective of the ongoing Pierre Auger Observatory~\cite{JM}, which will use both
techniques simultaneously, will be to resolve this discrepancy and to produce
a single coherent spectrum at the highest energies.

In the recent years, an interesting phenomenon has been noted by
Berezinsky and collaborators~\cite{BGG}. These authors have shown that
pair production losses of UHECR protons emitted by a cosmological
population of sources would modify
the energy spectrum in such a way as to mimic the ankle. The ankle 
would no longer mark the emergence of UHECRs out of the all-particle spectrum
but rather the signature of sources at cosmological
distances! Obviously a cut-off at $\epsilon_{\rm GZK}$ is then 
expected unless another (unspecified) source conspires to produce
trans-GZK particles. It seems reasonable to assume that counterparts
have not been detected in this case as they are too faint because too
far. Nonetheless one
may note  that intriguing correlations between catalogs of
UHECR arrival directions and catalogs of BL Lacertae sources have been
reported~\cite{TT}. The chance of coincidence is as low as
$10^{-3}\to10^{-5}$ depending on the catalogs and cuts chosen~\cite{TT};
however there is an ongoing debate on the actual significance of these 
correlations, 
see Ref.~\cite{TT2}.

  One may wonder where and how does the UHECR component emerge out of the
all-particle spectrum in this scenario. It turns out that it is
necessary for the extra-galactic UHECR spectrum to cut off somewhere
below $\sim10^{18}\,$eV, otherwise it would overproduce the
all-particle flux; in Ref.~\cite{BGG}, a cut-off on the UHECR
injection spectrum at this energy was postulated. Interestingly
another feature of the all-particle spectrum is seen at this energy,
namely the ``second knee'', where the differential spectrum 
steepens~\cite{NW00}.  Moreover, results of the HiRes experiment
suggest that protons dominate the chemical composition at energies 
$E\gtrsim 10^{18}\,$eV, and that heavy nuclei dominate below. 
It is then tempting to believe that the transition between the
Galactic and extra-galactic cosmic rays actually occurs around the 
second knee and not at the ankle, as previously thought.

  This scenario thus proposes the following interpretation of the
cosmic-ray spectrum. The ``first knee'' at
$E\sim2\cdot10^{15}\,$eV is generally believed to correspond to  the 
steepening of the Galactic proton spectrum, either because of a  change of 
propagation regime or because of a maximal energy limitation at the
source. Recent KASCADE data suggest that the region between
$\sim10^{15}\,$eV and $\sim10^{17}\,$eV is the superposition of the
knees of chemical species of increasing mass, with a knee position
scaling with the atomic number~\cite{KASCADE}; the knee for iron group
elements, in particular, is expected at $E\sim7\cdot10^{16}\,$eV. 
Finally the region between this iron
knee and the second knee would be the superposition of the iron (heavy)
Galactic component and of the emerging extra-galactic proton component,
in agreement with the HiRes measurements of chemical composition. In
this economical scenario, there is no need for a third source of
cosmic rays to account for the intermediate region between the iron
knee and the ankle, which is certainly attractive.

  This scenario requires to tune the low-energy cut-off of the
extragalactic spectrum to the high energy cut-off of the Galactic component.
Recently, it was suggested that the low-energy cut-off might not be a
result of injection processes but of propagation effects in
extra-magnetic fields~\cite{L05}. The basic mechanism is that
below some critical energy, particles take longer than the age of the 
Universe to travel from the source to the detector. Hence this latter
is actually shielded at low energies by a cosmological magnetic
horizon and the spectrum cuts off. In detail, 
the distance traveled in a Hubble time $H_0^{-1}$ by particles of
energy $E$ and scattering length $\lambda$ in extra-galactic
magnetic fields  is $d\sim \left(\lambda
H_0^{-1}\right)^{1/2}\sim 65\,{\rm Mpc}\,\lambda_{\rm Mpc}^{1/2}$
($H_0=70\,$km/s/Mpc), with $\lambda_{\rm Mpc}$ in units of Mpc. If the
typical UHECR source distance is $\sim 50\,$Mpc, as suggested by
statistical studies of clustering at the highest
energies~\cite{clusters}, a low energy cut-off appears for energies
such that $\lambda\ll 1\,$Mpc (the scattering
length is an increasing function of energy). For continuously emitting 
sources and a scattering length $\lambda\propto E^2$, a value of
$B\sqrt{l_{\rm c}}\sim 2\cdot 10^{-10}\,{\rm G}\cdot{\rm Mpc}^{1/2}$ 
produces a nice fit of the all-particle cosmic ray spectrum~\cite{L05}. 

  It is quite interesting to note that in this scenario, future
experiments such as KASCADE-Grande~\cite{KASCADE}, which will probe the
chemical composition and the spectrum around the second knee would
also probe the energy scaling of the scattering length, hence the
configuration and strength of extra-galactic magnetic fields. 
This scenario has been revisited recently and results in agreement
with those of~\cite{L05} were obtained~\cite{AB05}. For the range of
values of $B\sqrt{l_{\rm c}}$ which would explain the transition
between the Galactic and extra-galactic components, one predicts
deflection angles $\sim\,{\cal O}(1^o)$ at $E\sim10^{20}\,$eV, hence
the angular images of point sources could help probe the
extra-galactic magnetic field strength, and test this scenario.

\section{Large-scale magnetic fields.}

The previous discussion has brought in the possible importance of
extra-magnetic fields in high energy cosmic ray phenomenology. As a
matter of fact, as this section is to show, extra-galactic magnetic
fields are at the heart of the enigma of ultra-high energy cosmic
rays. With one or two exceptions, all astrophysical models for the
origin of UHECRs (i.e., those not invoking new
physics) require the existence of widespread magnetic fields to
explain the absence of counterparts in the arrival directions of the
highest energy events. One main unknown in this area of research remains
the strength and configuration of these magnetic fields.

Their existence has been asserted notably by the detection of synchrotron
emission bridging two components of a supercluster of
galaxies~\cite{Kea89}. Whether these fields are patchy, all pervading,
and what their amplitude is, remains however unknown; if at
equipartition with the ambient supercluster gas, one expects
$B\,\sim\,0.1\,\mu$G. Observations~\cite{KV} of the Faraday rotation of distant
polarized sources have revealed magnetic fields of strength $\sim\mu$G
in the heart of clusters of galaxies but yielded only upper limits on
the average extra-galactic $B\sqrt{l_{\rm c}}$. For a homogeneous
random magnetic field (meaning that $\vert \mathbf B\vert$ is
homogeneous,  and $\mathbf B$ random), 
this limit~\cite{W02} is $B\sqrt{l_{\rm c}}\lesssim
10^{-8}\,$G$\cdot$Mpc$^{1/2}$. Since magnetic fields tend to
follow the baryon density, it is reasonable to expect 
magnetic fields stronger in superclusters and filaments of large scale
structure than in voids. In these walls and filaments, 
the upper limit~\cite{RKB}
is $B\sqrt{l_{\rm c}}\lesssim 10^{-6}\,$G$\cdot$Mpc$^{1/2}$.

Theoretical predictions for $B$ and $l_{\rm c}$ are not reliable 
because the origin of the extra-galactic (and even of the Galactic) 
magnetic field is unknown. Sites proposed range from the
inflationary area to reionization to galactic pollution of the
intergalactic medium~\cite{W02}. In the end, one cannot predict reliably
the strength, coherence length and distribution of extra-galactic
magnetic fields to within orders of magnitudes.

Consequently, with regards to ultra-high energy cosmic ray
phenomenology, various cases have been considered. The present
discussion will consider in turn the two extreme cases: 
that of ``weak'' magnetic fields
$10^{-12}\,{\rm G}\cdot{\rm Mpc}^{1/2}\ll\,B\sqrt{l_{\rm
c}}\,\ll\,10^{-8}\,{\rm G}\cdot{\rm Mpc}^{1/2}$, and that of
``strong'' magnetic fields $10^{-8}\,{\rm G}\cdot{\rm
Mpc}^{1/2}\lesssim\,B\sqrt{l_{\rm c}}$.  Magnetic fields with strength
$B\sqrt{l_{\rm c}}\,\lesssim\,10^{-12}\,{\rm G}\cdot{\rm Mpc}^{1/2}$
are too weak to affect UHECR protons in a significant way.

\subsection{``Weak'' extra-galactic magnetic fields and bursting sources.}

For magnetic fields in the range $10^{-12}\,{\rm G}\cdot{\rm
Mpc}^{1/2}\ll\,B\sqrt{l_{\rm c}}\,\ll\,10^{-8}\,{\rm G}\cdot{\rm
Mpc}^{1/2}$, UHECRs with energy close to the GZK cut-off
$\epsilon_{\rm GZK}$ propagate quasi-rectilinearly, performing a random
walk with deflection angles $\delta\theta \ll 1$ at each crossing of a
cell of coherence of the magnetic field. The total rms deflection
angle for a source at distance $d$ reads~\cite{WM}:\\ $\theta_{\rm
rms}\,\simeq\, 2.5^\circ\,\left(E/ 10^{20}\,{\rm eV}\right)^{-1}
\left(d/ 100\,{\rm Mpc}\right)^{1/2} \left(B\sqrt{l_{\rm c}}/
10^{-9}\,{\rm G\cdot Mpc^{1/2}}\right)$. The associated time delay
with respect to straight line propagation can be expressed as:\\ 
$\tau\,\simeq\,
1.5\cdot10^5\,{\rm yr}\,\left(E/ 10^{20}\,{\rm eV}\right)^{-2}
\left(d/ 100\,{\rm Mpc}\right)^2 \left(B\sqrt{l_{\rm c}}/
10^{-9}\,{\rm G\cdot Mpc^{1/2}}\right)^{2}$.

If the time delay is greater than the characteristic timescale on which
the source is active, one cannot expect to see a counterpart. A
prototypical model is the $\gamma-$ray burst scenario of UHECR
origin~\cite{WV}. There the existence of magnetic fields is also
required to reconcile the observed $\gamma-$ray burst rate in the GZK
sphere with the detection rate of particles of energy
$\sim\epsilon_{\rm GZK}$. In effect, the former is $\sim
5\cdot10^{-4}$ $\gamma-$ray burst per year in a (GZK sphere) 
volume $(100\,{\rm Mpc})^3$, while the latter is
much higher, being $\sim 0.1-1$ UHECR with energy $\sim10^{20}\,$eV
per year. Since particles suffer stochastic energy losses and
propagate through stochastic magnetic fields, UHECRs of a given energy
arrive at any given point with a finite time spread
$\Delta\tau$. Provided this time dispersion is much larger than the
time interval between two consecutive $\gamma-$ray bursts in the GZK
sphere, the detector registers a ``continuous'' flux of
UHECRs. Assuming that $\Delta\tau \approx \tau$, one finds that
$B\sqrt{l_{\rm c}}\gtrsim 3\cdot10^{-10}\,{\rm G\cdot Mpc^{1/2}}$
suffices to reconcile these rates.

From the point of view of the detector, the $\gamma-$ray burst is seen
 in a limited band of energy~\cite{WM,Lea97} since $\tau\propto
E^{-2}$: at a given time one sees only UHECRs of a
given energy, others have passed by or are yet to pass. However one should not
expect to detect this correlation between arrival time and energy, since
the scatter $\Delta\tau$ is expected to be much larger than the
lifetime of an experiment. However, the energy width of the signal
depends on the ratio $\Delta\tau / \tau$ which in turn depends on the
configuration of the magnetic field. Hence there is hope to constrain
this latter with the collection of a significant number of UHECR events from
a single $\gamma-$ray burst~\cite{WM,SL}.

\subsection{``Strong'' localized magnetic fields.}

As mentioned previously, extra-galactic magnetic fields of strength
$B\sqrt{l_{\rm c}}\gtrsim 10^{-8}\,{\rm G}\cdot{\rm Mpc}^{1/2}$ must
be distributed inhomogeneously; they are likely to be enhanced in the
regions of high baryon density, i.e. walls and filaments of
large-scale structure. It then becomes mandatory to model the configuration
of these magnetic fields, which brings in more unknown and
unconstrained parameters. Although this area of research has been
quite active in the past years, it is probably fair to say that its
rich phenomenology has not been explored thoroughly yet. However, a
few trends and several main effects have been noted, as discussed
below.

  One major difference with the case of ``weak'' magnetic fields is
that UHECRs diffuse instead of wandering in a quasi-rectilinear
fashion~\cite{WW,BO,SLB,LSB,FP,LSC,AB04}. 
Hence the spectrum itself is affected by
propagation. Not all UHECRs diffuse however: those of the highest
energies suffer little angular deflection and quasi-rectilinear propagation
remains a good approximation. The critical energy at which this
separation of propagation regime occurs can be found by matching the
time of quasi-rectilinear propagation, accounting for the time delay
given above, and the diffusive propagation time $\tau_{\rm
diff}=d^2/(4D)$, where $d$ denotes the linear distance to the source and $D$
the diffusion coefficient. This latter is a function of energy that
depends on the configuration of the magnetic field. For instance, for
Komogorov turbulence $\langle B^2({\mathbf k})\rangle \propto
\vert{\mathbf k}\vert^{-11/3}$, one finds $D(E) \propto E^{1/3}$ for
$r_{\rm L}\ll l_{\rm c}$ and $D(E) \propto E^2$ for $r_{\rm L}\gg
l_{\rm c}$, with $r_{\rm L} = 1.1\,{\rm Mpc}\,(E/10^{20}\,{\rm
eV})(B/1\,\mu{\rm G})^{-1}$ the Larmor radius~\cite{CLP}. For this
case, one finds~\cite{BO,SLB,AB04} that particles with energy $E\ll
5\cdot10^{19}\,$eV diffuse if the source lies at linear distance $d\sim
10\,$Mpc and the magnetic field $B\sqrt{l_{\rm c}}\sim
0.1\,\mu$G$\cdot$Mpc$^{1/2}$. One general trend is that diffusion of 
sub-GZK energy particles tend
to soften the GZK cut-off. In effect diffusion steepens the spectrum 
at sub-GZK energies since the propagated differential spectrum $j(E)\propto
Q(E)/D(E)$ [$Q(E)$ injection spectrum] due to the increased
residence time. Hence the fit to the measured
flux requires a comparatively harder injection spectrum. Consequently 
the flux at trans-GZK energies, where the
spectrum is unaffected by the magnetic field, is comparatively higher.

  An important consequence of the presence of ``strong''
extra-galactic magnetic fields is the isotropization of the
arrival directions for UHECRs regardless of the source
distribution~\cite{WW,BO,LSB,FP,LSC,MT}. Obviously, this would explain the
absence of counterpart to the highest energy events. For instance, a
population of sources concentrated in the Local Supercluster, which is
seen in the sky as a band of $\sim10^\circ$ width, could produce an
isotropic UHECR sky as seen from Earth (to within present instrumental
sensitivity), provided $B\sim0.1-0.3\,\mu$G~\cite{LSB,FP,LSC}. 
Searches for anisotropy remain
important, as they might allow to constrain the source distribution
and the amount of magnetic bending~\cite{LSC}. In the same
models, it has been found that if the distance scale to the sources is
$\sim10\,$Mpc, one could also reproduce the rather hard energy
spectrum recorded by AGASA. A larger distance scale would result in a
more pronounced GZK cut-off, which depending on the resolution of the
HiRes-AGASA discrepancy, might be in better agreement with the data.
 
  Yet another effect is that of magnetic
lensing~\cite{WM,SL,SLB,LSB,HMR}. It has
been found that chaotic magnetic fields tend to focus and defocus the
trajectories of UHECRs in a process akin to gravitational lensing of
light rays. Interestingly, the doublets and triplets of events 
seen in the same direction by different experiments might  be magnetic 
mirages instead of pointing back to a point-like source~\cite{LSB}.  
Magnetic lensing is not restricted to strong turbulent magnetic
fields, but could also be produced by ``weak'' magnetic fields~\cite{WM,SL}
and by the Galactic magnetic field in the case of UHECR nuclei~\cite{HMR}. 
Sophisticated analytical methods borrowed
from gravitational lensing studies have been developed to reconstruct
the magnetic field from (hopefully future) observations of magnetic
lenses~\cite{lensing-reconstruct}.

  Previous considerations have assumed that the magnetic field is
purely random and the turbulence isotropic. However one cannot exclude
the possibility of a coherent component, for instance aligned along
the filament or the plane of the supercluster. This case implies
energy dependent anisotropy~\cite{regular}. This can be understood by the fact that
transport perpendicular to the coherent magnetic field is inhibited
with respect to longitudinal transport. Hence the exact spectrum as
seen by the detector depends on the orientation of the intervening
magnetic fields.

  Several studies have examined the propagation of heavy nuclei 
in extra-galactic magnetic fields~\cite{heavy}; in a first
approximation, one finds effects similar to those seen for protons,
albeit for a magnetic field strength rescaled by the atomic number.

  Some recent studies have considered the possible influence of
localized strong magnetic fields on the propagation of UHECRs, notably
in clusters of Galaxies~\cite{local}. However, the covering factor of
clusters of Galaxies is so small that  only those UHECRs
emitted by sources inside the clusters themselves are likely to be affected.  Along
similar lines, one should note the studies related to the Galactic
magnetic field and its influence on the propagation of
UHECRs~\cite{gal-wind,galactic}. In particular, one important aspect is how to
deconvolve the effect of the Galactic field from the data in order to
recover the extra-galactic signal~\cite{galactic}. 
The magnitude of the angular deflection suffered by
$10^{20}\,$eV protons coming from $b=0^o$ Galactic latitudes
is $\sim 5^o$, while at higher Galactic latitudes, $b=\pm45^o$, the deflection
falls to $\sim 1^o$.

  Finally, two groups have attempted to obtain definite predictions
for the effect of extra-galactic magnetic fields on UHECR propagation
by performing ab-initio MHD simulations of large-scale structure
formation, scaled to reproduce existing magnetic field data in
clusters of galaxies~\cite{SME,Dea}. The results obtained differ widely
between these two groups, however. One group~\cite{Dea} finds
typical deflections $\lesssim 1^o$ at $E\sim 4\cdot10^{19}\,$eV,
concluding that UHECRs should point back to
the sources. The other group~\cite{SME} finds that angular 
deflection is so large ($\gtrsim 50^o$) at the same energy that 
charged particle astronomy with
future experiments should not be possible. The
difference stems mostly from the assumptions made on the origin of the
magnetic field, and stands as a clear demonstration of how little is 
known about extra-galactic magnetic fields.

\section{Outlook.}

  Many questions have been raised and left unanswered, and the current
theoretical understanding of ultra-high energy cosmic ray phenomenology might
appear slightly confused. At which energy does the UHECR component
emerge out of the all-particle spectrum? Does the ankle mark this
transition? Does it rather say that UHECR sources
are located at cosmological distances? What is the effect of
extra-galactic magnetic fields on the spectrum and angular images of
ultra-high energy cosmic rays? Do the arrival directions of UHECRs
point back to the sources?

  This apparent confusion actually attests of the wealth and of the maturity of
this field of research. One may hope that future large-scale
experiments, notably the Pierre Auger Observatory, will answer the
above questions and show the way to the source of UHECRs.


\section*{References}
\begin{thebibliography}{99}

\bibitem{NW00} M. Nagano, A. A. Watson, \Journal{\RMP}{72}{689}{2000}.

\bibitem{gzk} K. Greisen, \Journal{\PRL}{16}{748}{1966}; 
Z.T. Zatsepin, V.A. Kuz'min, 
\Journal{\em Zh.Eksp.Teor.Fiz.Pis'ma Red.}{4}{144}{1966}

\bibitem{gzk-heavy} J.-L. Puget, F.W. Stecker, J.H. Bredekamp, 
\Journal{\ApJ}{205}{638}{1976}

\bibitem{BW} J. Bahcall, E. Waxman,
\Journal{\em Phys. Lett. B}{556}{1}{2003}

\bibitem{dMBO} D. de Marco, P. Blasi, A. Olinto, 
\Journal{\em Astropart. Phys.}{20}{53}{2003} 


\bibitem{HiRes} R. U. Abbasi {\it et al.} (HiRes collaboration), 
\Journal{\PRL}{92}{151101}{2004}.

\bibitem{AGASA} M. Takeda {\it et al.} (AGASA collaboration),
\Journal{\PRL}{81}{1163}{1998}; 
{\tt http://www-akeno.icrr.u-tokyo.ac.jp/AGASA/}

\bibitem{JM} J. Matthews, these proceedings.

\bibitem{BGG} V. Berezinsky, A. Gazizov, S. Grigorieva, 
{\tt arXiv:hep-ph/0204357}; {\tt arXiv:astro-ph/0210095}; V. Berezinsky, 
S. Grigorieva, B. Hnatyk, \Journal{\em Astropart. Phys.}{21}{617}{2004}.

\bibitem{TT} P.G. Tinyakov, I.I. Tkachev, \Journal{\em JETP Lett.}{74}{445}{2001}
[{\em Pisma Zh.Eksp.Teor.Fiz.}74:499 (2001)]; D.S. Gorbunov, P.G. Tinyakov, 
I.I. Tkachev, \Journal{\ApJ}{577}{L93}{2002}; P.G. Tinyakov, I.I. Tkachev,
 \Journal{\PRD}{69}{128301}{2004}; D.S. Gorbunov, these proceedings.

\bibitem{TT2} G. Sigl, D. Torres, L. Anchordoqui, G. Romero, 
\Journal{\PRD}{63}{081302}{2001}; D. Torres, S. Reucroft, O. Reimer, 
L. Anchordoqui, \Journal{\ApJ}{595}{L13}{2003}; N.W. Evans, F. Ferrer, 
S. Sarkar, \Journal{\PRD}{67}{103005}{2003}, \Journal{\PRD}{69}{128302}{2004}.

\bibitem{KASCADE} K.-H. Kampert {\it et al.} (KASCADE collaboration), 
{\tt arXiv:astro-ph/0405608}.

\bibitem{L05} M. Lemoine, {\tt arXiv:astro-ph/0411173}.

\bibitem{clusters} S.L. Dubovsky, P.G. Tinyakov, I.I. Tkachev, 
\Journal{\PRL}{85}{1154}{2000}; H. Yoshiguchi, S. Nagataki, 
S. Tsubaki, K. Sato, \Journal{\ApJ}{586}{1211}{2003}; P. Blasi, D. de Marco, 
\Journal{\em Astropart. Phys.}{20}{559}{2004}; 
M. Kachelriess, D. Semikoz, {\tt arXiv:astro-ph/0405258}.

\bibitem{AB05} R. Aloisio, V. Berezinsky, {\tt arXiv:astro-ph/0412578}.

\bibitem{Kea89} K.-T. Kim, P.P. Kronberg, G. Giovannini, T. Venturi, 
\Journal{\em Nature}{341}{720}{1989}

\bibitem{KV} P. Kronberg, \Journal{\em Rep. Prog. Phys.}{57}{325}{1994};
J. P. Vall\'ee, \Journal{\em New Astron. Rev.}{48}{763}{2004}

\bibitem{W02} L.M. Widrow, \Journal{\RMP}{74}{775}{2003}

\bibitem{RKB} D. Ryu, H. Kang, P.L. Biermann, \Journal{335}{19}{1998};
P. Blasi, S. Burles, A. Olinto, \Journal{\ApJ}{514}{L79}{1999}; 
G. Farrar, T. Piran, \Journal{\PRL}{84}{3527}{2000}

\bibitem{WM} E. Waxman, J. Miralda-Escud\'e, \Journal{\ApJ}{472}{L89}{1996}

\bibitem{WV} E. Waxman, \Journal{\PRL}{75}{386}{1995};
M. Vietri, \Journal{\ApJ}{453}{883}{1995}


\bibitem{Lea97} M. Lemoine, G. Sigl, A. Olinto, D. Schramm, 
\Journal{\ApJ}{486}{L115}{1997}; 
T. Stanev, R. Engel, A. Mucke, R. Protheroe, J. Rachen, 
\Journal{\PRD}{62}{093005}{2000}; 
A. Achterberg, Y. Gallant, C. Norman, D. Melrose, 
{\tt arXiv:astro-ph/9907060}.

\bibitem{SL} G. Sigl, M. Lemoine, \Journal{\em Astropart. Phys.}{9}{65}{1998}

\bibitem{WW} J. Wdowczyk, A. W. Wolfendale, \Journal{\em Nature}{281}{356}{1979};
M. Giler, J. Wdowczyk, A. W. Wolfendale, \Journal{\em J. Phys. G}{6}{1561}{1980};
V.S. Berezinsky, S.I. Grigorieva, V.A. Dogiel, 
\Journal{\em Sov. Phys. JETP}{69}{453}{1989} [{\em Zh. Eksp. Teor. Fiz.}96:798 (1989)]

\bibitem{BO} P. Blasi, A. Olinto, \Journal{\PRD}{59}{023001}{1999}

\bibitem{SLB} G. Sigl, M. Lemoine, P.L. Biermann, 
\Journal{\em Astropart. Phys.}{10}{141}{1999}

\bibitem{LSB} M. Lemoine, G. Sigl, P.L. Biermann, {\tt arXiv:astro-ph/9903124}

\bibitem{FP} G. Farrar, T. Piran, {\tt arXiv:astro-ph/0010370}; 
L. Anchordoqui, H. Goldberg, T.J. Weiler, \Journal{\PRL}{87}{081101}{2001};
C. Isola, M. Lemoine, G. Sigl, \Journal{\PRD}{65}{023004}{2002}

\bibitem{LSC} Y. Ide, S. Nagataki, S. Tsubaki, H. Yoshiguchi, K. Sato, 
\Journal{\em Pub. Astron. Soc. Japan}{53}{1153}{2001}; 
C. Isola, G. Sigl, \Journal{\PRD}{66}{083002}{2002}

\bibitem{AB04} R. Aloisio \& V. Berezinsky, {\tt arXiv:astro-ph/0403095}

\bibitem{CLP} F. Casse, M. Lemoine, G. Pelletier, 
\Journal{\PRD}{65}{023002}{2002}; 
J. Candia, E. Roulet, \Journal{\em JCAP}{0410}{007}{2004}

\bibitem{MT} G. Medina-Tanco, T. Ensslin, 
\Journal{\em Astropart. Phys.}{16}{47}{2001} 

\bibitem{HMR} D. Harari, S. Mollerach, E. Roulet, 
\Journal{\em JHEP}{9908}{022}{1999}, 
\Journal{\em JHEP}{0002}{035}{2000}, 
\Journal{\em JHEP}{0010}{047}{2000} 

\bibitem{lensing-reconstruct} D. Harari, S. Mollerach, E. Roulet, 
\Journal{\em JHEP}{0203}{045}{2002}
\Journal{\em JHEP}{0207}{006}{2002}

\bibitem{regular} G. Medina-Tanco, \Journal{\ApJ}{505}{L79}{1998};
T. Stanev, D. Seckel, R. Engel, \Journal{\PRD}{68}{103004}{2003}

\bibitem{heavy} L. Anchordoqui, H. Goldberg, S. Reucroft, J. Swain, 
\Journal{\PRD}{64}{123004}{2001}; G. Bertone, C. Isola, M. Lemoine, 
G. Sigl, \Journal{\PRD}{66}{103003}{2002}; G. Sigl, 
\Journal{\em JCAP}{0408}{012}{2004}

\bibitem{local} C. Rordorf, D. Grasso, K. Dolag, 
\Journal{\em Astropart. Phys.}{22}{167}{2004}; 
T. Wibig, A. Wolfendale, {\tt arXiv:astro-ph/0406511}

\bibitem{gal-wind} E.-J. Ahn, G. Medina-Tanco, P.L. Biermann, T. Stanev, 
{\tt arXiv:9911123}

\bibitem{galactic} T. Stanev, P. L. Biermann, J. Lloyd-Evans,
J. P. Rachen, A. A. Watson, \Journal{\PRL}{75}{3056}{1995}; T. Stanev, 
\Journal{\ApJ}{479}{290}{1997}; G. A. Medina-Tanco, E. M. de Gouveia
dal Pino, J. E. Horvath, \Journal{\ApJ}{492}{200}{1998}; D. Harari,
E. Mollerach, E. Roulet, \Journal{\JHEP}{08}{022}{1999}; 
J. Alvarez-Mu\~niz, R. Engel, T. Stanev, \Journal{\ApJ}{572}{185}{2002};
P.G. Tinyakov, I.I. Tkachev, \Journal{\em Astropart. Phys.}{18}{165}{2002};
M. Prouza, R. Smida, \Journal{\em Astron. Astrophys.}{410}{1}{2003};
H. Yoshiguchi, S. Nagataki, K. Sato, \Journal{\ApJ}{596}{1044}{2003}, 
\Journal{\ApJ}{607}{84}{2004};
P. G. Tinyakov, I. I. Tkachev, {\tt arXiv:astro-ph/0411669};


\bibitem{SME} G. Sigl, F. Miniati, T. Ensslin, 
\Journal{\PRD}{68}{043002}{2003}, 
\Journal{\PRD}{70}{043007}{2004}

\bibitem{Dea} K. Dolag, D. Grasso, V. Springel, I. Tkachev, 
\Journal{\em JETP Lett.}{79}{583}{2004} [{\em Pisma Zh.Eksp.Teor.Fiz.}79:719 (2004)];
{\tt arXiv:astro-ph/0410419}

\end{thebibliography}
\end{document}